\documentstyle[sprocl,psfig]{article}

\newcommand{\cpt}{\chi PT}

\newcommand{\veps}{{\vec\epsilon}}
\newcommand{\vepsprime}{{\vec\epsilon\, '}}

\begin{document}

\title{DEUTERON COMPTON SCATTERING IN CHIRAL PERTURBATION THEORY}

\author{M. MALHEIRO}

\address{Instituto de F\'\i sica, Universidade Federal Fluminense,
24210-340, Niter\'oi, Brazil}

\author{S.R. BEANE, D.R. PHILLIPS}

\address{Department of Physics, University of Washington,
Seattle, WA 98195-1560}

\author {U. VAN KOLCK}

\address{Department of Physics, University of Arizona, Tucson, AZ 85721\\
RIKEN-BNL Research Center, Brookhaven National Laboratory,
Upton, NY 11973\\
Kellogg Radiation Laboratory,
Caltech, Pasadena, CA 91125}

\maketitle

\abstracts
{Compton scattering on the deuteron is studied in the framework of
baryon chiral perturbation theory to third order in small momenta, for
photon energies of order the pion mass. The scattering amplitude is a
sum of one- and two-nucleon mechanisms with no undetermined
parameters.  Our results are in good agreement with the 
intermediate energy experimental data, and a comparison is made with the
recent higher-energy data obtained at SAL.}

\section{Introduction}
\label{sec-intro}

At sufficiently low energy $\omega$ the spin-averaged forward Compton
scattering amplitude for any nucleus is, in the nuclear rest frame:

\begin{equation}
T(\omega)=-\frac{1}{4\pi} \vepsprime \cdot \veps
\left(\frac{{\cal Z}^2 e^2}{A M} + (\alpha + \beta) \omega^2 + \ldots \right),
\end{equation}
where $\veps$ and $\vepsprime$ are the polarization vectors of the
initial and final-state photons, ${\cal Z} e$ is the total charge of
the nucleus, $A M$ is its total mass, and $\alpha$ ($beta$) the electric
(magnetic) polarizability. 

Compton scattering on a deuteron target is an obvious candidate to measure 
the electromagnetic neutron polarizabilities. 
In this case, the cross section in
the forward direction roughly goes as

\begin{equation}
\left.\frac{d \sigma}{d \Omega} \right|_{\theta=0}
\sim (f_{Th} - (\alpha_p + \alpha_n) \omega^2)^2.
\end{equation} 
The sum $\alpha_p + \alpha_n$ may then be accessible via its
interference with the dominant Thomson term for the proton,
$f_{Th}$. 
Coherent scattering the deuteron has been measured at $E_\gamma=$ 49
and 69 MeV by the Illinois group \cite{lucas}.  An experiment with
tagged photons in the energy range $E_\gamma= 84.2-104.5$ MeV  at Saskatoon
has recently presented results \cite{SAL}, while data for $E_\gamma$ of about
60 MeV is being analyzed at Lund.

\begin{figure}[t]
\psfig{figure=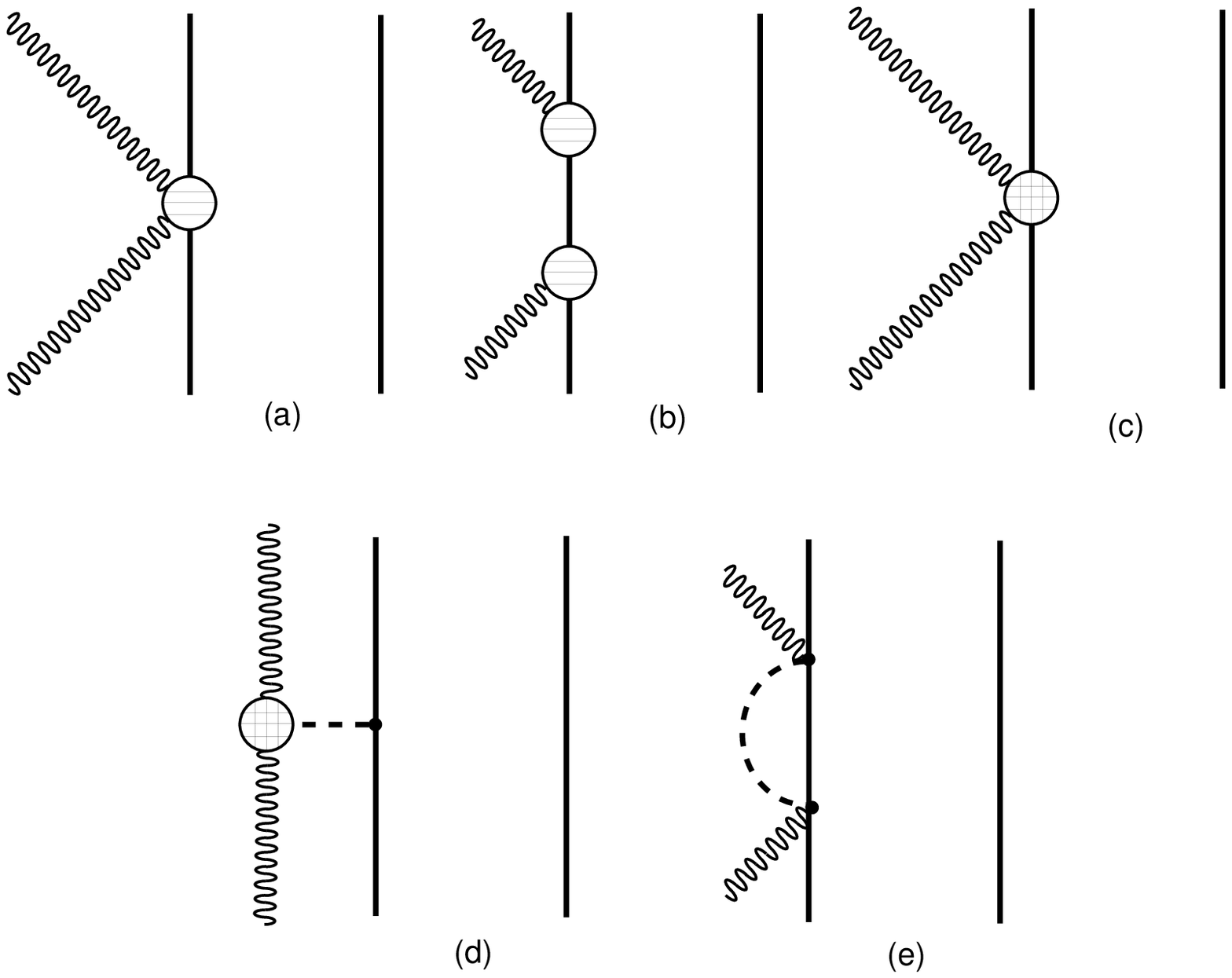,height=6cm,width=5.5cm}
\vspace*{-6cm}
\hspace*{3cm}
\centerline{\psfig{figure=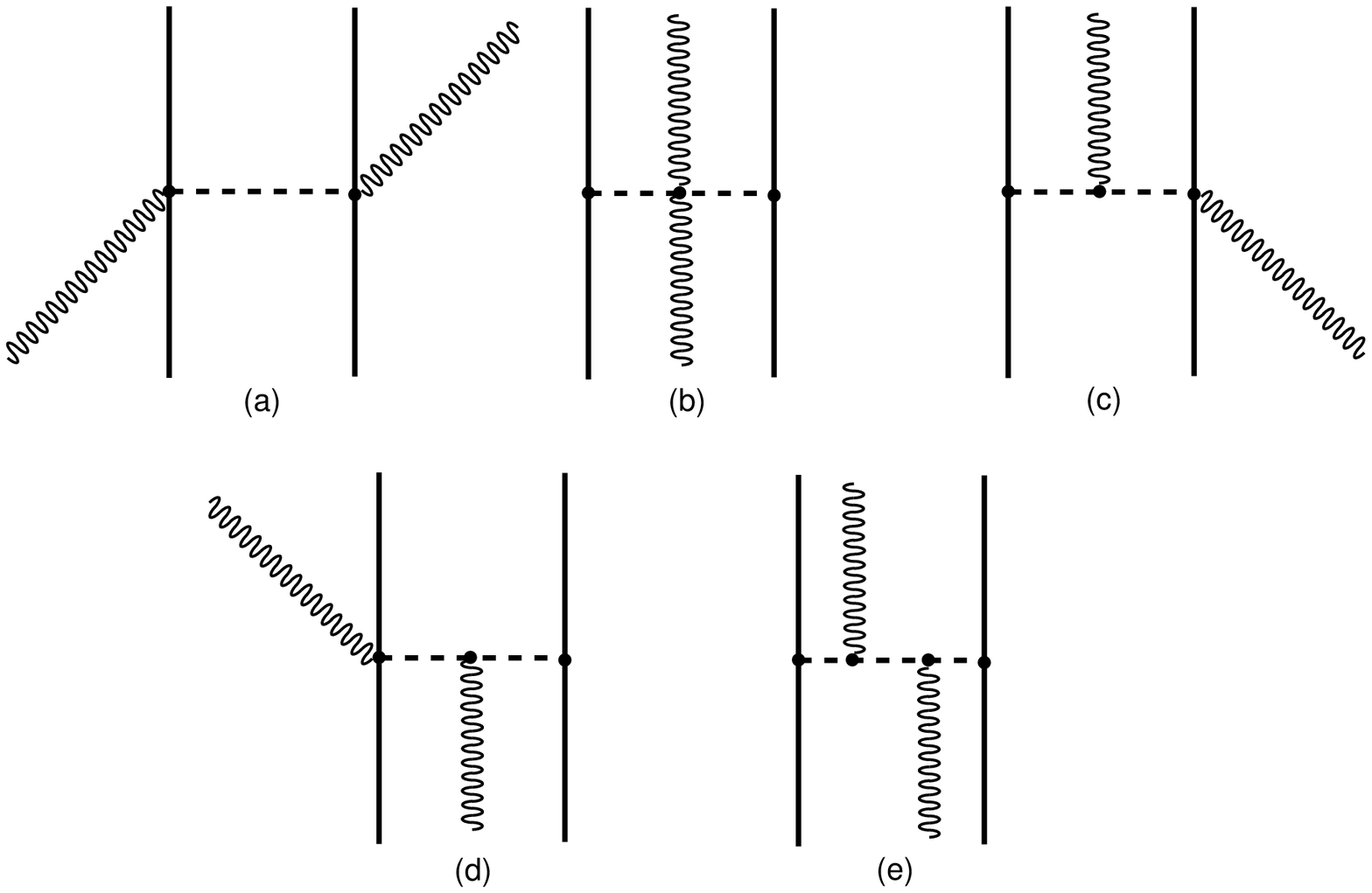,height=6cm,width=5.5cm}}
\parbox{5.5cm}{\caption{\label{fig1} Characteristic
   one-body interactions which contribute to Compton scattering on the
   deuteron at order $Q^2$ (a) and at order $Q^3$ (b-e) (in the
   Coulomb gauge). The sliced blobs are vertex insertions from
   ${\cal L}^{(1)}$. The sliced and diced blobs are vertex insertions
   from ${\cal L}^{(2)}$.}}
\hspace*{0.75cm}
\parbox{5.5cm}{\caption{\label{fig2}
Two-body interactions which contribute to Compton scattering
on the deuteron at order $Q^3$. Permutations are not shown.}}
\end{figure}

\section{The $\cpt$ calculation}
\label{sec-calc}

In this section, we outline how the various contributions are
calculated following Ref.\cite{weinnp}. We can separate 
them into two classes: one-nucleon and two-nucleon mechanisms. 
One-nucleon diagrams are those which relate directly to
contributions to Compton scattering on a single nucleon, while
two-nucleon mechanisms contribute only in $A\ge 2$ nuclei. 

Let us consider first the one-nucleon terms; a sample of the relevant
diagrams is shown in Fig. \ref{fig1}.  
The only modification which must be made to the photon-nucleon
amplitude for use in Compton scattering on the deuteron
is that it must be boosted from the $\gamma N$ c.m. frame to the
$\gamma NN$ c.m. frame.

In principle we cannot hope to extract neutron polarizabilities without a
theory for the two-nucleon contributions,
shown in Fig. \ref{fig2}.  Comparison of
diagrams in Figs. \ref{fig1}(e) and \ref{fig2} shows that the same
physics that produces dominant contributions to the polarizabilities 
generates the dominant two-nucleon contributions. 

\begin{figure}[t,h,b,p]
\psfig{figure=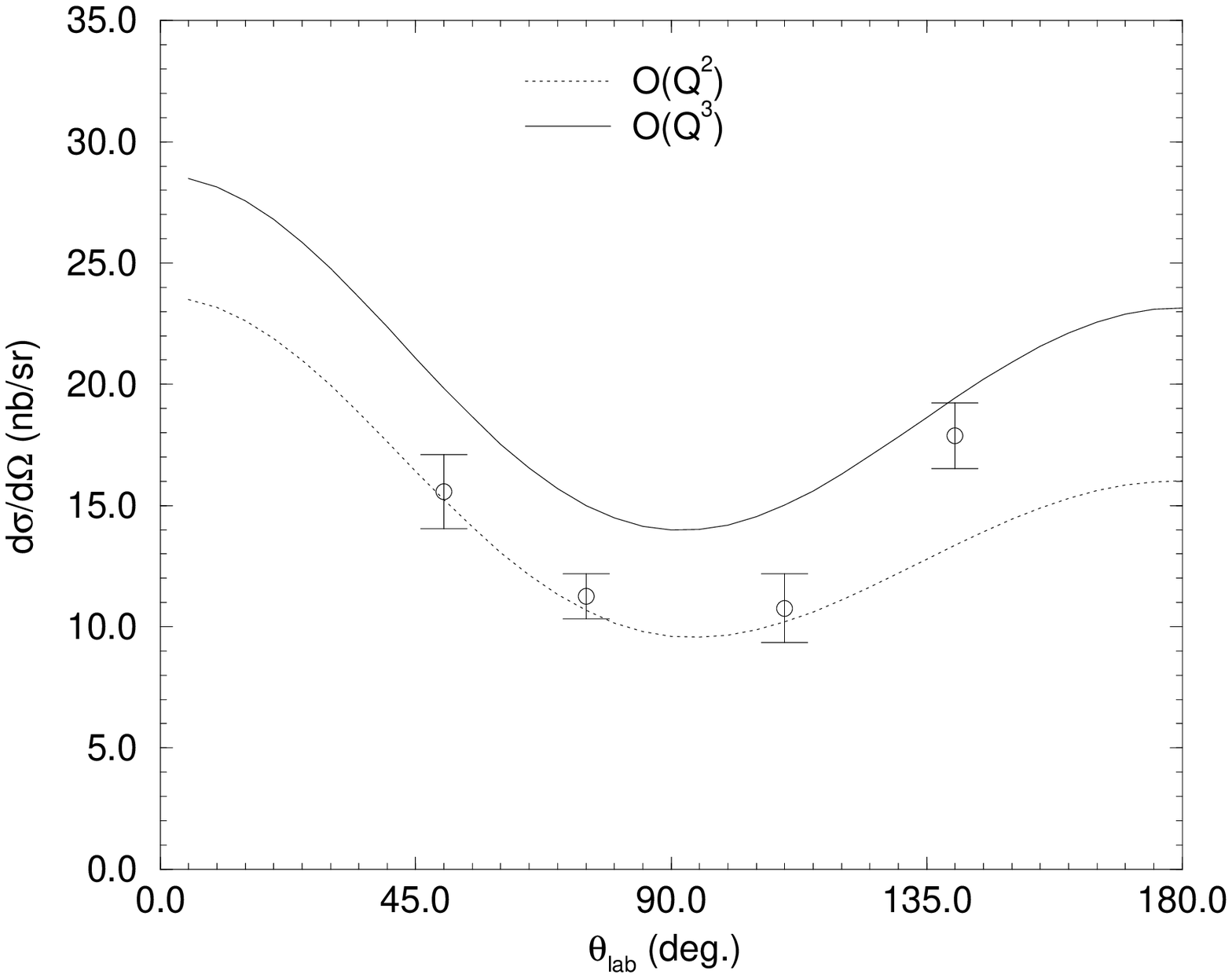,height=6cm,width=6cm}
\vspace*{-6cm}
\hspace*{3cm}
\centerline{\psfig{figure=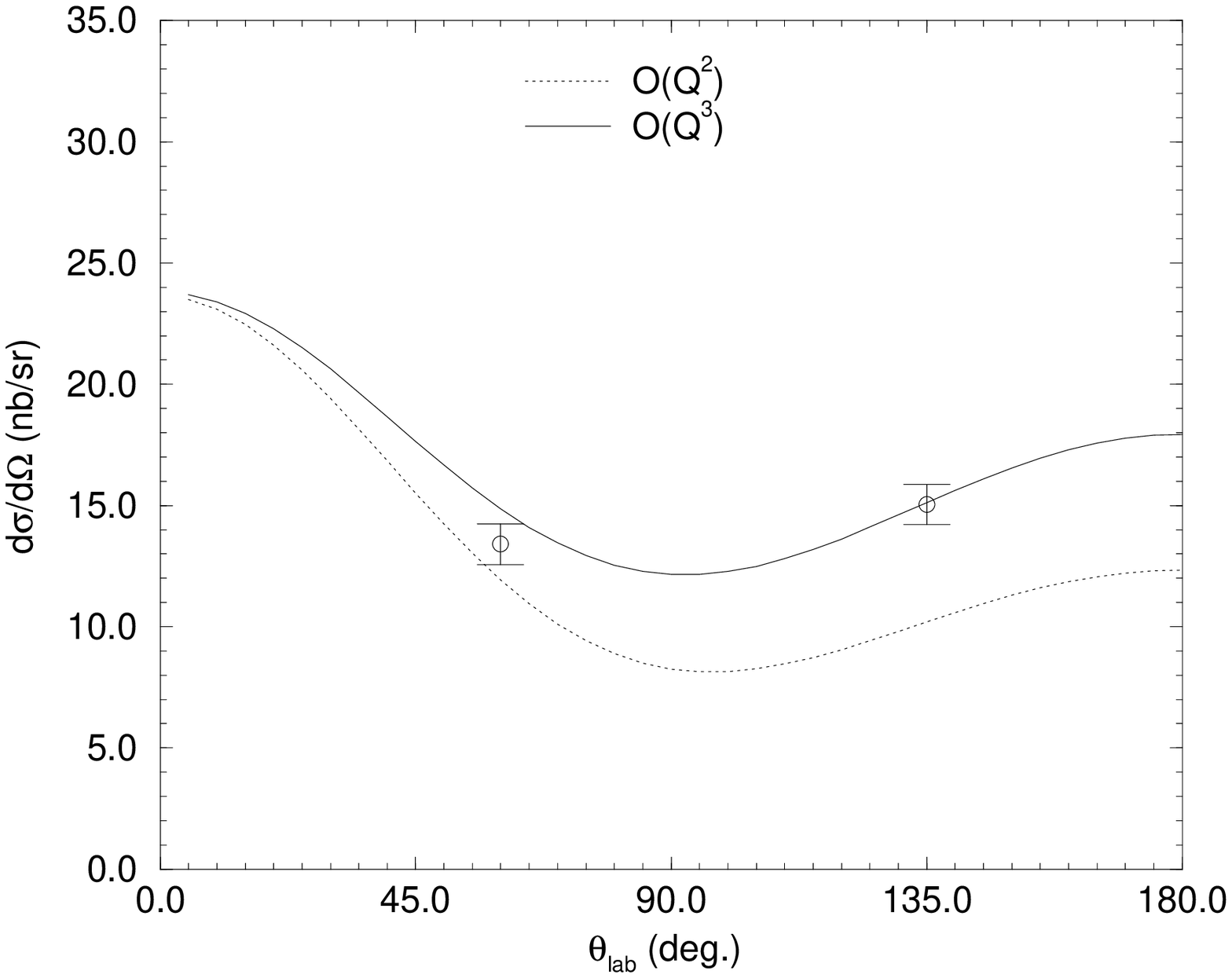,height=6cm,width=6.5cm}}
\parbox{5.5cm}{\caption{\label{fig6} Results of the
   $O(Q^2)$ (dotted line) and $O(Q^3)$ (solid line) calculations
   at a photon laboratory energy of 49 MeV. The data points of
   Ref.~\protect\cite{lucas} are also shown.}}
\hspace*{0.75cm}
\parbox{5.5cm}{\caption{\label{fig7} Results of the
   $O(Q^2)$ (dotted line) and $O(Q^3)$ (solid line) calculations
   at a photon laboratory energy of 69 MeV. The data points of
   Ref.~\protect\cite{lucas} are also shown.}}
\end{figure}

\begin{figure}[t,h,b,p]
\label{fig4}
\centerline{\psfig{figure=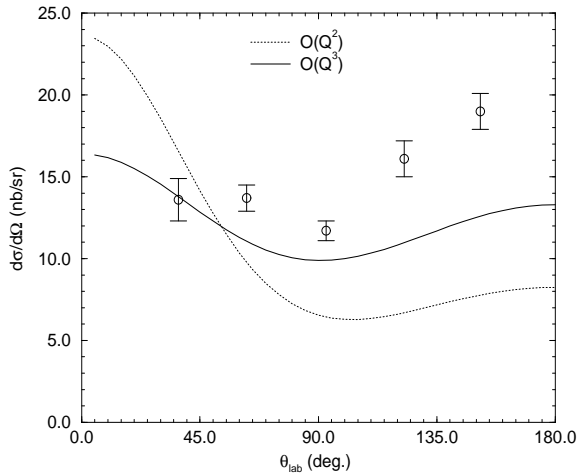,width=3.0in,height=2.5in}}
\caption{\label{fig8} Results of the
   $O(Q^2)$ (dotted line) and $O(Q^3)$ (solid line) calculations
   at a photon laboratory energy of 95 MeV. The recent SAL data
points of Ref.~\protect\cite{SAL} are also shown.}
\end{figure}

We calculate the cross section for Compton scattering on the deuteron
including the single-scattering and two-nucleon mechanisms described
above.  
The final result is insensitive to details of the deuteron
wavefunction, and below
we use the energy-independent
Bonn OBEPQ wave function.  The photon-deuteron $T$-matrix 
is then calculated and the laboratory differential cross section
evaluated directly from it:

\begin{equation}
  \frac{d \sigma}{d \Omega_L}=\frac{1}{16 \pi^2}
  \left(\frac{E_\gamma'}{E_\gamma}\right)^2 \frac{1}{6} \sum_{M'
    \lambda' M \lambda} |T^{\gamma d}_{M' \lambda' M \lambda}|^2,
\end{equation}
where $E_\gamma$ and $E_\gamma'$ are the initial and final photon energy 
in the laboratory frame respectively.
Results are shown in Figs. \ref{fig6},\ref{fig7},\ref{fig8}.
They represent the 
$\cpt$ {\it predictions} for Compton scattering on the
deuteron \cite{silas}.


\section{Conclusion}
\label{sec-conclusion}

We have calculated the differential cross section
for Compton scattering on the deuteron in $\cpt$ up to
$O(Q^3)$~\cite{silas}. We have found:

$\bullet$ Reasonable agreement with the data at 49 MeV.  At this
energy $O(Q^3)$ corrections are not large compared to the leading
$O(Q^2)$ result.  

$\bullet$ Good agreement with the data at 69 MeV.  At this energy the
convergence appears to be good. This suggests that $\cpt$ at $O(Q^3)$ is
providing reasonable neutron and two-nucleon contributions.

$\bullet$ A prediction at 95 MeV which is, however, plagued by
considerable uncertainties.  The cross
section comes out somewhat smaller than at lower energies, in
particular in the backward directions, although the full $O(Q^4)$
amplitude is likely to be somewhat bigger at back angles. It
seems that a more stringent test of $\cpt$ at these energies will have to
wait for a next-order calculation.



\section*{References}
\begin{thebibliography}{99}

\bibitem{lucas}
M.~Lucas, Ph.~D. thesis, University of Illinois, unpublished (1994). 

\bibitem{SAL} D.L. Hornidge et al., Phys. Rev. Lett.{\bf 84}, 2334 
(2000). 

\bibitem{weinnp}  S. Weinberg, Phys. Lett. {\bf B251}, 288 (1990); 
              Nucl. Phys. {\bf B363}, 3 (1991);
              Phys. Lett. {\bf B295}, 114 (1992); 
C. Ord\'{o}\~{n}ez and U. van Kolck, Phys. Lett. {\bf B291}, 459
(1992); U. van Kolck, Phys. Rev. {\bf C49}, 2932 (1994).

\bibitem{silas} S.R.~Beane, M.~Malheiro, D.R.~Phillips, and U. van
Kolck, Nucl. Phys. {\bf A656}, 367 (1999).

\end{thebibliography}
\end{document}